\documentclass[prb,aps,showpacs,preprint]{revtex4}
\usepackage{epsfig}
\begin{document}
\title{Suppression  of inelastic bound state resonance effects by the dimensionality of atom-surface scattering event}
\author{A. \v{S}iber and B. Gumhalter}
\affiliation{Institute of Physics, P.O. Box 304,
10001 Zagreb, Croatia}

\date{\today}
\begin{abstract}
We develop a multidimensional coupled channel method suitable for studying the interplay of bound state resonance and phonon assisted scattering of inert gas atoms from solid surfaces in one, two and three dimensions.  This enables us to get insight into the features that depend on the dimensionality of inelastic resonant processes typically encountered in low energy He atom scattering from surfaces, in general, and to elaborate on the observability of recently conjectured near threshold resonances in scattering from Einstein phonons, in particular.   
\end{abstract}


\maketitle

\newcommand{\bq}{\begin{equation}}
\newcommand{\eq}{\end{equation}}

\newcommand{\barr}{\begin{eqnarray}}
\newcommand{\earr}{\end{eqnarray}}

A unified treatment of inelastic, resonant and diffractive scattering of inert gas atoms from solid surfaces has remained a challenging task since the very beginning of development of quantum atom-surface scattering theory in the 1930s\cite{GoodmanRev}. Already at that stage it was noted that a class of processes involving emission and absorption of phonons combined with the intermediate projectile propagation through bound states of the static atom-surface potential may take place and thereby modify the scattering amplitudes. However, in the works quoted in Ref. \onlinecite{GoodmanRev} the descriptions of inelastic scattering and accommodation were limited to one-phonon processes that were treated essentially within the one-dimensional (1D) framework, i.e. only the component of the projectile momentum normal to the surface was assumed to change upon phonon emission or absorption. The three dimensional (3D) treatment was implemented only in the description of diffraction and selective adsorption processes.    

Further developments of the early quantum atom-surface scattering theories proceeded mainly in three directions: (i) full 3D unitary treatment of diffraction of atoms from surfaces\cite{CCGM}, (ii) full 3D unitary treatment of one-phonon inelastic scattering\cite{CM} and its extension by inclusion of diffraction\cite{GoodmanTan,Eich}, and (iii) quantum description of sticking and accommodation processes\cite{BrivioGrimley}, dominantly within the one-phonon theories\cite{Brenigstick}, with extension to the selective adsorption processes\cite{Boeheim}. A more complex problem of multiphonon atom-surface scattering, important for instance for the determination of ubiquitous Debye-Waller factor (DWF) and pronounced satellite structures in the scattering spectra, was tackled a bit later and mainly within the semiclassical theories\cite{BortoLevi,BK}. Only recently a tractable fully quantum scattering spectrum formalism for treating multiphonon atom-surface scattering has been developed\cite{plrep} and applied to the interpretation of inelastic He atom scattering (HAS) spectra for the various systems in which the diffractive and inelastic spectral features could be easily disentangled. Although the scattering spectrum  formalism offers exact solutions also in the case of a strong interplay between multiphonon inelastic and elastic resonant processes (i.e. diffraction and selective adsorption) its applicability in this situation is hindered by computational difficulties.  

Diffractive scattering of atoms from ordered solid surfaces can be efficiently treated in the absence of inelastic processes by the method of coupled channel (CC) equations\cite{Wolken,LinWolken,SecrestJohnson,Hutson}.  
Application of the CC method to the calculations of diffraction spectra for He atoms scattered from Xe monolayer adsorbed on graphite\cite{Hutson,ComKKK} demonstrated the potentiality of the method in the treatment of surface scattering. Partial inclusion of the effect of substrate phonons through the Debye-Waller attenuations of diffraction peak intensities\cite{SibGumXegraphite} then posed a question of the extension of the CC approach so as to enable the assessment of resonant and inelastic effects and their interplay on an equal footing. This is possible in principle but difficulties arise in the implementation of such a demanding numerical algorithm to 3D scattering in which the phonon configuration space is multi-dimensional\cite{LinWolken}. On the other hand, the treatment of 1D inelastic scattering by Einstein phonons (colinear motion of the projectile and surface oscillator) by the CC method poses no difficulties\cite{SecrestJohnson} and it was first applied in atom-surface scattering  to investigate the Debye-Waller factors\cite{Kasai} and the long standing problem of the importance of linear and nonlinear projectile-phonon coupling in inelastic scattering in the absence of bound states\cite{nonlinscatt}. Extension of this analysis by the inclusion of bound state resonances\cite{BrenigPRL,BreGum} showed that the latter can indeed significantly affect the amplitudes of 1D low energy scattering near the single-phonon excitation threshold, in a sequence of strongly correlated events of single- and multiphonon  excitations and intermediate projectile propagation through one or more bound states. These findings have posed a fundamental question as to whether such a practically exact unified 1D treatment of resonant and inelastic scattering can predict and interpret\cite{BrenigPRL} some new features in the 3D high resolution-low energy angular resolved HAS from surfaces\cite{Toenniesrev}.

The answer to the above posed question obviously requires extension of the CC method to description of inelastic scattering beyond 1D. This is almost a formidable task if the phonon configuration space is multi-dimensional, i.e. if the phonon quantum numbers (wavevector and polarization) span a multi-dimensional Fock configuration space. 
To overcome this difficulty we introduce here a tractable restricted Fock space coupled channel (RFCC) algorithm for treating inelastic atom-surface scattering involving bound state resonances that applies equally well to 1-, 2- and 3D collisions. 

A prerequisite for accurate description of low-energy inelastic atom-surface scattering is a correct treatment of the momentum transfer that takes place between the projectile and substrate phonons. To implement this requirement in a numerical treatment we begin in a standard fashion and discretize the phonon wave vector space\cite{LinWolken}. We take $N$ discrete values $( {\bf Q}_1,{\bf Q}_2,...,{\bf Q}_N)$ of phonon wavevector ${\bf Q}$ that are  
distributed homogeneously over the first surface Brillouin
zone. If one takes into 
account the states with $j$ phonons excited, the dimension of the basis set describing the 
subset of these states grows as $N^j$. Owing to this, a numerical treatment of higher excited phonon states $j>2$ for $D>1$, even with a sparsely discretized space of phonon wave vectors, is beyond the capabilities of the present day  computers. Hence, in the present RFCC approach we consider only the phonon ground-state, the subset of $N$ singly excited states, {\em and} the subset of $N(N-1)/2$ doubly excited states, i.e. our restricted Fock space of excited states contains  $(N^2+N+2)/2$ different phonon states that should suffice for describing typical regimes of HAS experiments carried out to reveal the characteristics of surface phonons in the single phonon excitation regime\cite{Toenniesrev} and beyond\cite{VAS,HeXePRL,HeXePRB}. 

We start from the total wave function of the system

\begin{equation}
\Psi ({\bf R},z, \{{\bf u}_l \}) = \sum_{n} \phi_{n}(z) \exp (-i {\bf K}_{n} {\bf R}) \otimes |n \rangle
\label{eq:ansatz2}
\end{equation}
where ${\bf r} = ({\bf R}, z)$ is the radiusvector of the projectile (of mass $m$), with ${\bf R}$ lying in the surface plane,
the coordinates $\{{\bf u}_l \}$ describe the displacements of the target atoms of mass $M$, $l$ is the lattice site index, ${\bf K}_n$ is the projectile wavevector parallel to the surface, and $|n \rangle$ is a Fock state describing phonon excitations in the target. The functions $\phi_n (z)$ satisfy the CC equations

\begin{equation}
\frac{d^2}{dz^2} \phi_n (z) = \sum_{n'} W_{n,n'}(z) \phi_{n'} (z),
\label{eq:cc1}
\end{equation}
and obey the scattering boundary conditions\cite{SibGumXegraphite}. Henceforth we shall restrict the coupling to $z$-polarized displacements only and assume normal projectile incidence to enable equivalent treatment of the scattering event in all dimensions. The CC matrix elements are given by

\begin{equation}
W_{n,n'}(z) = \frac{2m}{\hbar ^2} V^{stat}(z)  \delta_{n,n'} - \kappa_{n}^2 \delta_{n,n'} 
+ \frac{2m}{\hbar^2} V^{dyn}_{n,n'}(z).
\end{equation}
The reduced channel energy, $\kappa_{n}^2$, is

\begin{equation}
\kappa_{n}^2 = \frac{2m}{\hbar ^2}(E_{i} - E_n) - K_{n}^2 ,
\label{eq:harmenerg}
\end{equation}
where $E_{i}$ is the projectile incident energy, $E_n =\sum_{p=1}^{N}  n_{{\bf Q}_p}  \hbar \omega({\bf Q}_p)$ is the vibrational energy of the target in the state $|n \rangle$ with 
$n_{{\bf Q}_p}$ denoting the occupation of ${\bf Q}_p$th phonon mode in the state $|n \rangle$, and $\omega({\bf Q})$ is the phonon frequency. Following the common practice we take the projectile-target atom binary potentials in separable form\cite{BortoLevi}  
$v({\bf r}-{\bf r}_l)=v_{D}(z-z_l)\exp\left[-({\bf R}-{\bf R}_l)^2/(2\sigma^2)\right]$, which when pairwise summed over $l$ yield for the average total static interaction potential:

\begin{equation}
V^{stat}(z) = \left(\frac{\sqrt{2 \pi} \sigma}{a}\right)^{D-1} v_{D}(z-z_0),\hspace{2mm}(D=1,2,3)
\label{eq:potstat}
\end{equation}
where $a$ is the nearest neighbor distance in the simple Bravais lattice of the target surface.
The matrix of the dynamic part of the interaction potential is given by

\begin{equation}
V^{dyn}_{n,n'}(z) = 
\left \{
\begin{tabular}{c}
$\frac{\partial V^{stat}(z)}{\partial z}  
\exp\left(-\frac{\sigma ^2{\bf Q}_r^2}{2}\right) 
\sqrt{\frac{\hbar(n_{{\bf Q}_r}+1)}{2MN \omega({\bf Q}_r)}};$
\\
$\exists r, 
n'_{{\bf Q}_r} = n_{{\bf Q}_r} + 1 , n_{{\bf Q}_q} = n'_{{\bf Q}_q}, \forall q \ne r$
\\
$\frac{\partial V^{stat}(z)}{\partial z} 
\exp\left(-\frac{\sigma ^2{\bf Q}_r^2}{2}\right) 
\sqrt{\frac{\hbar n_{{\bf Q}_r}}{2MN \omega({\bf Q}_r)}}; $
\\ 
$\exists r,
n'_{{\bf Q}_r} = n_{{\bf Q}_r} - 1 , n_{{\bf Q}_q} = n'_{{\bf Q}_q}, \forall q \ne r$
\\
0; \hspace{0.8mm} otherwise \\
\end{tabular}
\right. 
\label{eq:potdyn}
\end{equation}
 For computational convenience we have assumed only linear coupling of the projectile to the target atom displacements as this gives the dominant contribution to inelastic scattering\cite{nonlinscatt}. The above equations do not incorporate transitions involving the reciprocal lattice vectors ${\bf G}\ne 0$ i.e. they 
are strictly applicable to the case of smooth surfaces with $2\pi \sigma \gg a$ for which the diffraction probabilities are negligible. The resulting set of CC equations is then solved using standard techniques \cite{Wolken,LinWolken,SecrestJohnson,Hutson,ComKKK,SibGumXegraphite}.

\begin{figure}[tb]
\epsfxsize=8.5cm \epsffile{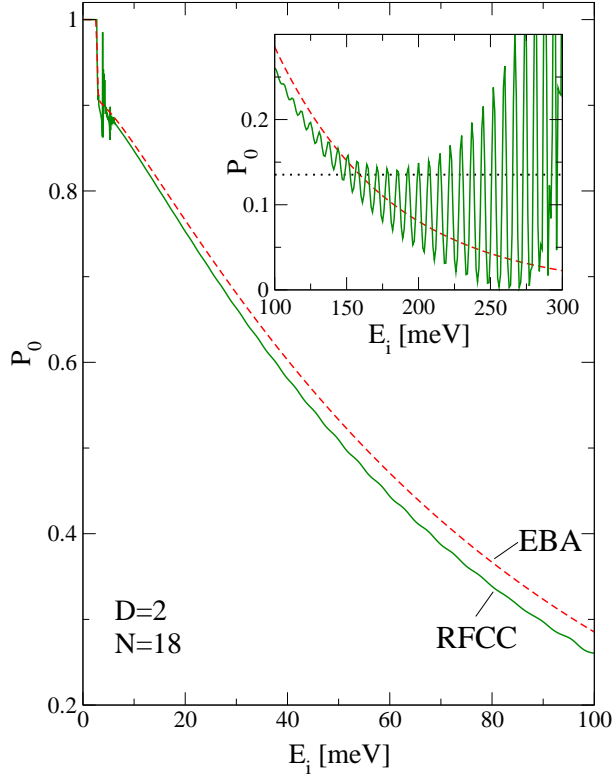}
\caption{Comparison of {\em two-phonon} RFCC and {\em multiphonon} EBA results for the Debye-Waller factor  $P_{0}$ in 2D He atom-surface scattering as a function of normal incident energy $E_{i}$. Model parameters are: $\hbar\omega_{0}=2.7$ meV, $\sigma=a=2$ \AA. }
\label{InelresFg1}
\end{figure}

\begin{figure}[tb]
\epsfxsize=8.5cm \epsffile{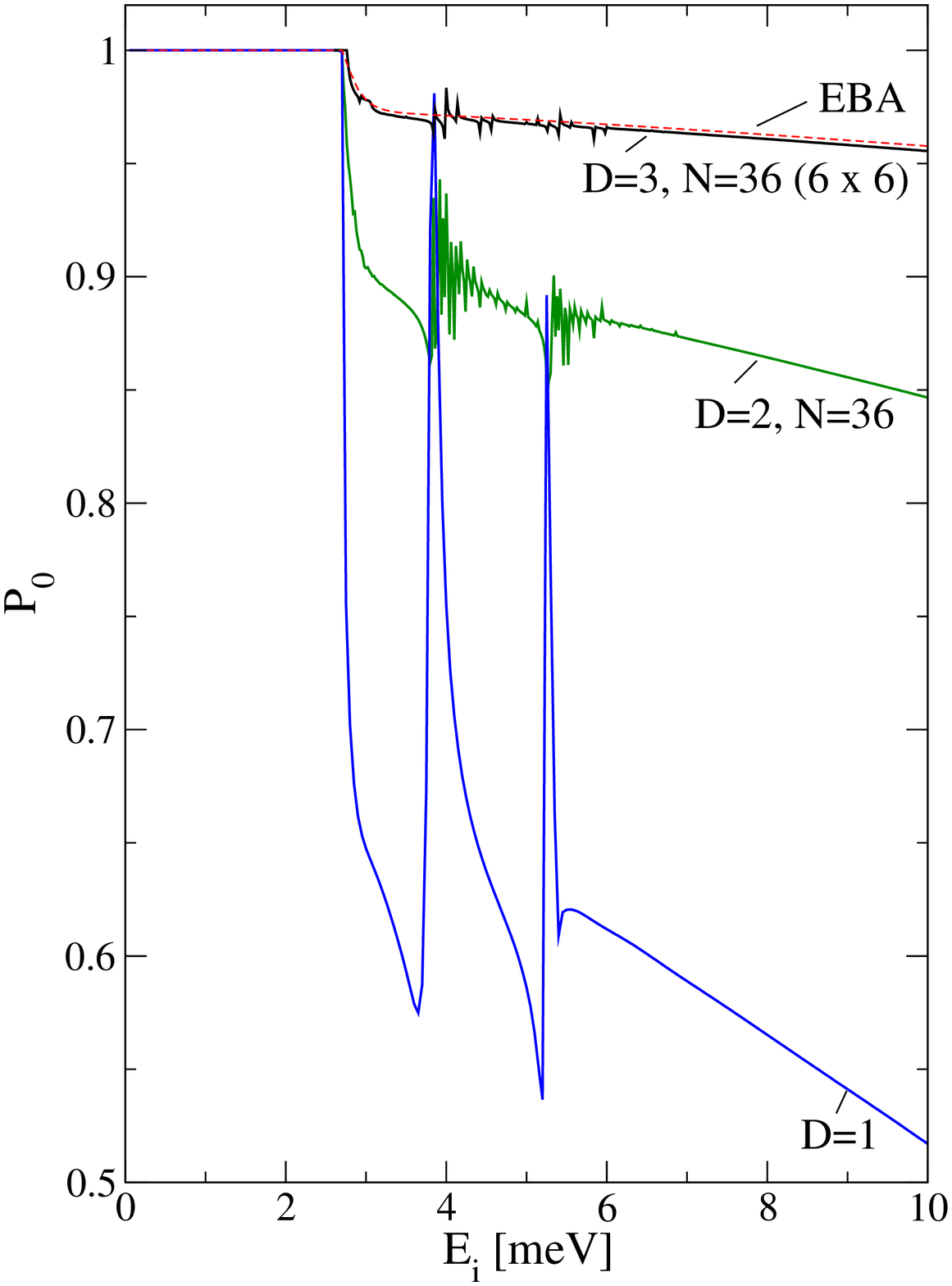}
\caption{Comparison of the RFCC Debye-Waller factors $P_{0}$ arising from interplay of Einstein phonon exchange and bound state resonance effects in 1D, 2D and 3D, and the $P_{0}$ obtained in 3D EBA (dashed line) for multiphonon He atom scattering. The resonances appear around $E_{i}=\epsilon_{1} +2\hbar\omega_{0}=3.83$ meV and $E_{i}=\epsilon_{2} +2\hbar\omega_{0}=5.3$ meV.}
\label{InelresFg2}
\end{figure}

The above outlined RFCC formalism is applied to the paradigmatic case of He atom scattering from a vibrating Xe monolayer adsorbed on a cold Cu(111) surface\cite{HeXePRL,HeXePRB,recovery,kinfoc}. The manifestations of inelastic bound state resonance effects are expected to be most pronounced in this prototype system because {\em (i)} the corrugation of the static He-Xe/Cu(111) potential gives rise to extremely weak diffraction peaks and hence the  ${\bf G}$-assisted transitions can be neglected, and {\em (ii)} among the three Xe-monolayer localized phonon modes the vertically polarized S-mode, which couples most strongly to the scattered He atom, exhibits no dispersion practically over the entire surface Brillouin zone, i.e. $\hbar\omega({\bf Q})=\hbar\omega_{0}=2.7$ meV. Hence, we can restrict our RFCC-analyses of the bound state resonance effects to the experimentally relevant situation in which He atoms are scattered by the vertically polarized Einstein phonons that propagate along a smooth surface.

 We represent the binary potential $v_{D} (z-z_0)$ by a Morse form, and adjust its well depth to obtain the same static component $V^{stat}(z)=V_{0}\left(\exp(-2z/b)-2\exp(-z/b)\right)$ of the 
interaction potential for all $D$. This gives rise to the transitions between the same energy levels so that the corresponding probabilities can be directly compared. The Morse parameters of $V^{stat}(z)$ are the same as in Refs. \onlinecite{BrenigPRL} and \onlinecite{BreGum}, i.e. $V_{0}=6.6$ meV and $b=0.82$ \AA, yielding $\epsilon_{0}=-4.54$, $\epsilon_{1}=-1.67$, and $\epsilon_{2}=-0.14$ meV. This choice facilitates comparison of the earlier calculated 1D results with the present ones. Note, however, that our 1D results can not be {\em identical} to those obtained in Refs. \onlinecite{BrenigPRL} and \onlinecite{BreGum} because in the present study the phonon Fock space is restricted and the model does not incorporate nonlinear coupling of the projectile to surface vibrations. As these restrictions are not easily removed in higher dimensions, we retain them here also in the 1D calculation to enable consistent comparison of the results obtained for different  dimensions.

The reliability of the present {\em two phonon} RFCC approach is tested by comparing the elastic scattering probabilities $P_{0}$ (i.e. DWF's) calculated by this method with the ones obtained from the {\em multiphonon} exponentiated Born approximation (EBA)\cite{plrep,BGL} that is exact in the low and high incident energy limits for all $D$ and produces accurate off-resonance results also at intermediate energies\cite{BGL,nonlinscatt,BrenigPRL,BreGum}. Figure  \ref{InelresFg1} compares $P_{0}$ obtained from the RFCC and EBA calculations for the intermediate case $D=2$. 
It should be noted that numerical requirements on these two types of calculations are quite different. The EBA formalism requires calculation of ${\bf Q}$-resolved single-phonon excitation probabilities and their integration over the first Brillouin zone\cite{plrep}. On the other hand, the RFCC calculations involve numerical 
propagation of large matrices [whose number of rows and columns is $(N^2+N+2)/2$] 
along the $z$-coordinate, and are thus much slower from the computational aspect (for the 3D case illustrated in Fig. \ref{InelresFg2} they are slower by a factor $\sim 10^{5}$). The displayed EBA and RFCC results significantly differ only in the low energy region where $P_{0}$ is affected by the resonant projectile propagation through the bound states by emission and reabsorption of phonons.
Inset shows the region of high incident energies in which the validity of the 2D RFCC calculation breaks down due to the neglect of Fock states with more than two excitation quanta. In the EBA, the average number of phonons $\bar{n}$ excited in a scattering event is related to $P_0$ through $P_0 = \exp (-\bar{n})$ \cite{plrep}. Thus, the displayed two phonon RFCC results are expected to be reliable in the scattering regime in which $P_0 > \exp (-2) = 0.135$, i.e. above the dotted line in the inset. The EBA predicts that this regime is reached for $E_{i}\sim$150 meV, which is precisely where the RFCC results begin to exhibit pronounced oscillatory  behavior and deviation from the corresponding asymptotically exact multiphonon EBA values. 
Hence, an excellent agreement between the results for off-resonant scattering in the range $0\leq E_{i} \leq 150$ meV obtained from two completely different algorithms confirms the validity of the present two-phonon RFCC approach to study the {\em interplay} of {\em multiphonon} and {\em resonant} scattering at low $E_{i}$.  
This is crucial because the resonant scattering effects predicted in the RFCC approach cannot be encompassed by the simplest version of the EBA even in the 1D case\cite{BrenigPRL,BreGum}.

The thus established validity of the two phonon RFCC for treating the interplay of multiphonon and resonant effects in the experimentally relevant  case of He atom scattering from Einstein phonons\cite{HeXePRL,HeXePRB} enables the assessment of the effect of dimensionality on these processes.
Figure \ref{InelresFg2} displays a comparison of the elastic scattering probabilities $P_{0}$ 
for He atoms incident normal to the "surface" of a 1D, 2D and 3D system. 
These results demonstrate that the inelasticity of 
scattering and the bound state resonance effects are reduced as $D$ increases. In the 1D case of colinear scattering the on-the-energy and momentum-shell requirements allow only the inelastic transitions for which ${\bf Q}=0$, and hence they are all governed by the maximum values of $V^{dyn}$ (cf. Eq. \ref{eq:potdyn}). On the other hand, with the increase of $D$ the on-shell scattering intensities are redistributed over the parts of momentum space with nonzero ${\bf Q}$ in which the magnitude of the matrix elements (\ref{eq:potdyn}) is reduced. As a result, as $D$ increases the total contribution to inelastic and resonant scattering intensities diminishes. Hence, although 
the signature of inelastic bound-state resonances persists in higher dimensions, it is already too weak in 3D to be experimentally observed for smooth surfaces on which the ${\bf G}$-assisted transitions are negligible.

In summary, we have developed a restricted Fock space coupled channels method suitable for a unified treatment of the interplay between multiphonon excitation and bound state resonance effects in atom-surface scattering. Application of the method to a prototype system of He atom scattering from Einstein phonons on smooth surfaces of adsorbed Xe monolayers shows that, in contrast to the earlier conjectures based on 1D models, such effects are only weakly discernible even under the most favorable experimental conditions for their observation\cite{kinfoc}.


\end{document}